\documentclass[amstex,prl,aps,twocolumn,psfig,showpacs]{revtex4}

\usepackage{amsmath}
\usepackage{amssymb}
\usepackage{graphicx}

\begin{document}

\title{First- and Second-Order Phase Transitions, Fulde-Ferrel Inhomogeneous
State and Quantum Criticality in Ferromagnet/Superconductor Double Tunnel
Junctions}
\author{Biao Jin, Gang Su$^{\ast}$ and Qing-Rong Zheng }
\affiliation{College of Physical Sciences, Graduate School of the Chinese Academy of
Sciences, P.O. Box 3908, Beijing 100039, China}

\begin{abstract}
First- and second-order phase transitions, Fulde-Ferrel (FF) inhomogeneous
superconducting (SC) state and quantum criticality in
ferromagnet/superconductor/ferromagnet double tunnel junctions are
investigated. For the antiparallel alignment of magnetizations, it is shown
that a first-order phase transition from the homogeneous BCS state to the
inhomogeneous FF state occurs at a certain bias voltage $V^{\ast }$; while
the transitions from the BCS state and the FF state to the normal state at $%
V_{c}$ are of the second-order. A phase diagram for the central
superconductor is presented. In addition, a quantum critical point (QCP), $%
V_{QCP}$, is identified. It is uncovered that near the QCP, the SC gap, the
chemical potential shift induced by the spin accumulation, and the
difference of free energies between the SC and normal states vanish as $%
|V-V_{QCP}|^{z\nu }$ with the quantum critical exponents $z\nu =1/2$, $1$
and $2$, respectively. The tunnel conductance and magnetoresistance are also
discussed.
\end{abstract}

\maketitle

\textit{Introduction.} ---Spin-dependent transport plays an essential role
in magnetic hybrid nanostructures in the field of spintronics (see, e.g.
Refs. \cite{meser,gijs,prinz,mood,wolf,sarma,book} for review). Among
others, ferromagnet/superconductor (F/S) heterostructures have attracted
much attention theoretically\cite%
{dejong,takahashi,jedema,zheng,tserko,yoshida,jin} and experimentally\cite%
{vasko,soulen,kontos,chen,gu,johansson,johnson} in recent years. For F/S/F
double tunnel junctions, it has been observed that the superconductivity is
suppressed by the injection of spin-polarized current (e.g. Refs. \cite%
{takahashi,zheng,tserko,yoshida,chen,jin,johansson}), that is due to the
nonequilibrium spin accumulation. When spin-polarized electrons are injected
into the superconductor, a spin density is accumulated near the interfaces
owing to the spin imbalance, thereby giving rise to an equivalent, small
magnetic field that acts as a pair-breaking field, which leads to a
suppression of superconductivity. There has been a recent study\cite%
{takahashi} showing that the \textit{homogeneous} superconducting (SC) state
is strongly suppressed with increasing the bias voltage and completely
destroyed at a critical voltage by the nonequilibrium spin density in the
antiparallel alignment of magnetizations. This study is qualitatively
consistent with the experimental observation for high biases, but is
inconsistent for low biases\cite{chen}.

On the other hand, about forty years ago, Fulde and Ferrel (FF)\cite{FF},
and Larkin and Ovchinnikov (LO)\cite{LO} independently, found that the SC
order parameter can be modulated in real space by a spin-exchange field of a
ferromagnet. Later, such an \textit{inhomogeneous} superconducting state has
been extensively explored under various circumstances (e.g. \cite%
{casal,kontos}). As the nonequilibrium spin accumulation may lead to an
equivalent magnetic field in the central supercondutor, the FFLO state,
which is simply omitted in the previous treatment\cite{takahashi}, might be
inevitable in the F/S/F double tunnel junction.

To understand profoundly the spin-dependent transport properties, in this
Letter, the F/S/F double tunnel junction shall be systematically revisited.
It is shown that in the antiparallel aligment, a first-order phase
transition from a homogeneous Bardeen-Cooper-Schrieffer (BCS) SC state to
the inhomogeneous FF SC state occurs at a certain bias voltage, while the
transitions from the BCS state and the FF state to the normal state are of
the second-order. A phase diagram for the central superconductor is
identified. Besides, a quantum critical point (QCP) is specified at bias $%
V_{QCP}$, near which the SC order parameter, the chemical potential shift
induced by the spin accumulation, and the difference of free energies
between the SC and normal states vanish as $|V-V_{QCP}|^{z\nu }$ with the
quantum critical exponents $z\nu =1/2$, $1$ and $2$, respectively. The
tunnel conductance and magnetoresistance are also obtained.

\textit{Model.} ---Consider a symmetric F/S/F double tunnel junction with
the left and right ferromagnetic (FM) electrodes applied by bias voltages $%
-V/2$ and $V/2$, respectively. The two identical FM electrodes are separated
from the central superconductor by two insulating thin films. The central
superconductor is presumably described within the framework of BCS theory.
Suppose that the energy relaxation time of quasiparticles is shorter than
the tunneling time, while the latter is shorter than the spin relaxation
time. As the resistance of this tunnel junction with insulating thin films
is greater than that of a conventional metallic contact, the Andreev
reflection effect can be reasonably ignored for simplicity.

From the standard tunneling Hamiltonian, and in light of the linear response
theory, the tunneling current through the $j$th junction can be readily
obtained by 
\begin{equation}
I_{j\sigma }=2\pi e\left\vert \widetilde{T}\right\vert ^{2}D_{j\sigma
}[N-\eta _{j}(\sigma S+\frac{Q-\widetilde{N}}{2})],  \label{current}
\end{equation}%
where$\ \widetilde{T}$\ is the tunneling matrix element,$\ j=1$, $2$, $%
D_{j\sigma }\ $is the subband density of states (DOS) in the $j$th FM
electrode, $\sigma =\pm 1$ for spin up and down, respectively, and $\eta
_{1}=1$, $\eta _{2}=-1$. The quantities $S$, $Q$, $N$\ and $\widetilde{N}$\
are defined by

\begin{eqnarray}
S &=&\frac{1}{2}\sum_{\mathbf{k}}(f_{\mathbf{k\uparrow }}-f_{\mathbf{%
k\downarrow }}),  \label{spin} \\
Q &=&\frac{1}{2}\sum_{\mathbf{k}}(\text{ }u_{\mathbf{k}}^{2}-v_{\mathbf{k}%
}^{2}\text{ })(f_{\mathbf{k\uparrow }}+f_{\mathbf{k\downarrow }}),
\label{charge} \\
N &=&\frac{1}{2}\sum_{\mathbf{k}}[f_{0}(E_{\mathbf{k}}-\frac{eV}{2}%
)-f_{0}(E_{\mathbf{k}}+\frac{eV}{2})],  \label{num-1} \\
\widetilde{N} &=&\frac{1}{2}\sum_{\mathbf{k}}(u_{\mathbf{k}}^{2}-v_{\mathbf{k%
}}^{2})[f_{0}(E_{\mathbf{k}}-\frac{eV}{2})+f_{0}(E_{\mathbf{k}}+\frac{eV}{2}%
)],  \label{num-t}
\end{eqnarray}%
where $f_{0}(z)$\ denotes the Fermi distribution function of thermal
equilibrium in FM electrodes, $f_{\mathbf{k}\sigma }$ is the nonequilibrium
distribution function of quasiparticles with energy $E_{\mathbf{k}}$\ and
spin $\sigma $ ($=\uparrow ,\downarrow $)\ in the central superconductor, $S$%
\ and $Q$\ represent the spin density and the quasiparticle charge density,
describing the spin imbalance and quasiparticle charge imbalance in the
central superconductor, respectively.

\textit{Fulde-Ferrel State.} ---As there appears the nonequilibrium spin
accumulation near the interfaces of the tunnel junction, it is believed that
the SC state would generally include the inhomogeneous FFLO state\cite{FF,LO}
in addition to the homogeneous BCS state. Without loss of generality, we
suppose that the SC order parameter takes the form of FF type\cite{FF}: $%
\Delta (\mathbf{r})=\Delta _{\mathbf{q}}e^{i\mathbf{q}\cdot \mathbf{r}}$
with $\Delta _{\mathbf{q}}$ the amplitude of the order parameter. When $%
\mathbf{q}=0$, it recovers the homogeneous BCS state. The quasiparticle
dispersion $E_{\mathbf{k}}$,\textrm{\ }the coherence factors $u_{\mathbf{k}}$%
\ and $v_{\mathbf{k}}$\ are given by $E_{\mathbf{k}}=\sqrt{\xi _{\mathbf{k}%
}^{2}+\Delta _{\mathbf{q}}^{2}}+(v_{F}q/2)x$, $u_{\mathbf{k}}^{2}=\frac{1}{2}%
(1+\xi _{\mathbf{k}}/\sqrt{\xi _{\mathbf{k}}^{2}+\Delta _{\mathbf{q}}^{2}})$%
, and $v_{\mathbf{k}}^{2}=\frac{1}{2}(1-\xi _{\mathbf{k}}/\sqrt{\xi _{%
\mathbf{k}}^{2}+\Delta _{\mathbf{q}}^{2}})$,\ respectively, where $\xi _{%
\mathbf{k}}$\ is the free electron energy relative to the chemical
potential, $v_{F}$\ the Fermi velocity and $x=\mathbf{k}\cdot \mathbf{q}%
/(qk) $ the cosine of the angle between $\mathbf{k}$ and momentum $\mathbf{q}
$ of a Cooper pair. The amplitude $\Delta _{\mathbf{q}}$\ is determined by
the gap equation%
\begin{equation}
1=\frac{V_{BCS}}{2}\sum_{\mathbf{k}}\frac{(1-f_{\mathbf{k\uparrow }}-f_{%
\mathbf{k\downarrow }})}{\sqrt{\xi _{\mathbf{k}}^{2}+\Delta _{\mathbf{q}}^{2}%
}},  \label{gap-eq}
\end{equation}%
where $V_{BCS}$\ is the BCS-type pair interaction. The values of $\mathbf{q}$
will be specified later.

Let us proceed to determine the nonequilibrium distribution function $f_{%
\mathbf{k}\sigma }^{F(A)}$. In the absence of spin-flip scattering, the spin
up and down tunneling currents are independent, and should be conserved,
i.e. $I_{1\sigma }=I_{2\sigma }$,\ yielding%
\begin{eqnarray}
S^{F} &=&0\text{ \ \ for the parallel alignment,}  \label{S-F} \\
S^{A} &=&PN^{A}\text{ \ \ for the antiparallel alignment,}  \label{S-A} \\
Q^{F} &=&Q^{A}=0\text{ \ \ for both alignments,}  \label{Q-FA}
\end{eqnarray}%
where the superscripts $F$\ and $A$\ refer to the parallel and antiparallel
alignments, respectively, and $P=\left\vert D_{j\uparrow }-D_{j\downarrow
}\right\vert /(D_{j\uparrow }+D_{j\downarrow })$\ is the spin polarization
of the FM electrodes. These solutions show that the nonequilibrium spin
accumulation exists \textit{only} in the antiparallel configuration. In the
above derivation, we have adopted the conventional constant DOS
approximation, $\sum_{\mathbf{k}}(\cdots )\simeq N(0)\int_{-\varpi
_{D}}^{\varpi _{D}}(\cdots )d\xi _{\mathbf{k}}$, where $\varpi _{D}$ is the
cut-off (Debye) energy, and $N(0)$ denotes the DOS of free electrons at the
Fermi level. The quantity $\widetilde{N}$ vanishes identically for both
alignments of magnetizations since the integrand is an odd function of $\xi
_{\mathbf{k}}$. Eq. (\ref{S-A}) requires that $f_{\mathbf{k\uparrow }}^{A}$
should differ from $f_{\mathbf{k\downarrow }}^{A}$ in the presence of the
tunneling current. Following Ref.\cite{takahashi}, we consider the solutions
of the form:%
\begin{eqnarray}
f_{\mathbf{k\sigma }}^{A} &=&f_{0}(E_{\mathbf{k}}-\sigma \delta \mu ),
\label{fA-updown} \\
f_{\mathbf{k\uparrow }}^{F} &=&f_{\mathbf{k\downarrow }}^{F}=f_{0}(E_{%
\mathbf{k}}),  \label{fF}
\end{eqnarray}%
where $\delta \mu $\ is introduced as the chemical potential shift induced
by the nonequilibrium spin accumulation, and plays essentially the same role
as the spin-exchange field explored by FF in their seminal article\cite{FF}.
This kind of solutions may be applicable if the thickness of the central
superconductor is much smaller than the spin diffusion length, and the spin
relaxation time is sufficiently long. Eqs. (\ref{fA-updown}) and (\ref{fF})
are the solutions of Eqs. (\ref{S-F}) and (\ref{Q-FA}). However, Eqs. (\ref%
{S-A}) and (\ref{gap-eq}) with Eqs. (\ref{fA-updown}) should be solved in a
self-consistent manner to specify $\delta \mu $ and $\Delta _{\mathbf{q}}$
as functions of the bias $V$, temperature $T$ and polarization $P$.

For these coupled equations, when the self-consistent multiple solutions
corresponding to different $\mathbf{q}$ appear, only the value of $\mathbf{q}
$ that leads to the lowest free energy of the system is retained. \textit{In
an inhomogeneous superconducting state, the nonzero solution for }$\Delta _{%
\mathbf{q}}$\textit{\ implies only the local minimum of the free energy,
which does not necessarily mean the stable state.} In order to clarify this
issue, one must compare the free energies of the homogeneous BCS, the
inhomogeneous FF and normal states, as emphasized by Abrikosov\cite%
{abrikosov}. The free energy of the present system can be obtained by
integrating the gap equation\cite{heslinga}%
\begin{eqnarray}
F_{S}^{F(A)}-F_{N}^{F(A)} &=&(\Delta _{\mathbf{q}}^{F(A)})^{2}/V_{BCS} 
\notag \\
&-&\int_{0}^{\Delta _{\mathbf{q}}^{F(A)}}dz\sum_{\mathbf{k}}(z/\sqrt{\xi _{%
\mathbf{k}}^{2}+z^{2}})  \notag \\
&\times &[1-f_{0}(\sqrt{\xi _{\mathbf{k}}^{2}+z^{2}}+\frac{v_{F}(\mathbf{%
q\cdot k)}}{2k}-\delta \mu ^{F(A)})  \notag \\
&-&f_{0}(\sqrt{\xi _{\mathbf{k}}^{2}+z^{2}}+\frac{v_{F}(\mathbf{q\cdot k)}}{%
2k}+\delta \mu ^{F(A)})],  \label{free-e}
\end{eqnarray}%
where $F_{S}^{F(A)}$and $F_{N}^{F(A)}$stand for the free energy of the SC
state and the N state, respectively, and $\delta \mu ^{F}=0$, $\delta \mu
^{A}=\delta \mu $. It turns out that both the homogeneous BCS ($\mathbf{q}=0$%
) and the FF ($\mathbf{q}\neq 0$)\ SC solutions are possible in the
antiparallel alignment, while the former solution is always favorable in the
parallel configuration.

\textit{Results.} ---In the parallel alignment of magnetizations, since
there is no spin and charge accumulation in this circumstance, the SC order
parameter does not depend on the bias voltage; while in the antiparallel
configuration, the situation becomes complicated, as the nonequilibrium spin
accumulation characterized by $S^{A}$ intervenes in. Figure 1 presents the
bias voltage dependence of the SC order parameter and the chemical potential
shift at $T/T_{c}=0.2$, with $T_{c}$ the SC critical temperature, for the
antiparallel alignment. It is observed that the order parameter $\Delta ^{A}$
remains almost constant at low biases and is in the homogeneous BCS state
till a specific bias voltage $V^{\ast }=1.36\Delta _{0}/e$ at which $\Delta
^{A}$ drops suddenly, where $P=0.4$, and $\Delta _{0}$ is the BCS
zero-temperature energy gap. Then, $\Delta ^{A}$ goes into the inhomogeneous
FF state with $\mathbf{q}\neq 0$, decreases with the bias, and vanishes
completely at $V=V_{c}$ where superconductivity is quenched, i.e. $\Delta _{%
\mathbf{q}}(V_{c})=0$, as shown in Fig. 1(a). At $V=V^{\ast }$, there is a
discontinuity for $\Delta ^{A}$, implying that a first-order phase
transition from the homogeneous BCS phase to the FF phase exists in the
system. The present result is quite different from that given in Ref.\cite%
{takahashi} where the inhomogeneous FF state was simply ignored. The
chemical potential shift grows in the homogeneous BCS state with increasing
the bias, and exbihits a jump at $V^{\ast }$, then increases slowly in the
FF state till $V_{c}$, and is linear for $V\geqslant V_{c}$ in the normal
state, as shown in Fig. 1(b). We have found that the free energy of the
inhomogeneous FF phase is lower than the homogeneous BCS phase for $%
V_{c}>V\geqslant V^{\ast }$, suggesting that the FF state is stable. At $%
V=V^{\ast }$, the free energy of the FF state coincides with that of the
homogeneous BCS state, revealing that the FF state can coexist with the BCS
state at $V^{\ast }$. 
\begin{figure}[tb]
\vspace{-0.85cm}
\begin{center}
\leavevmode\includegraphics[width=0.90\linewidth,clip]{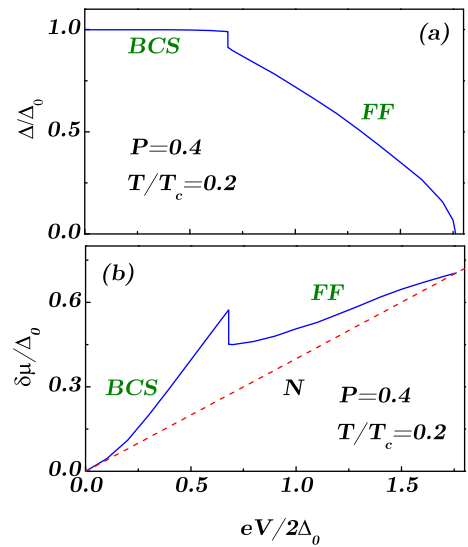}
\end{center}
\vspace{-1.0cm}
\caption{(Color online) The bias dependence of the order parameter $\Delta
^{A}$ (a) and the chemical potential shift (b) in the antiparallel alignment
of magnetizations at $P=0.4$, $T/T_{c}=0.2$, where a discontinuity from the
first-order phase transition is observed. The dashed line is for the normal
state.}
\label{fig1}
\end{figure}

The magnitude of momentum $\mathbf{q}$ of a Cooper pair is in general a
function of bias voltage, as shown in Fig. 2. At $V<V^{\ast }$, the system
is in the homogeneous BCS state, and thus $|\mathbf{q|}$ is zero; while for $%
V^{\ast }\leqslant V<V_{c}$, $|\mathbf{q|}$ corresponding to the lowest free
energy varies nonmonotonically with the bias, implying that the momenta of
Cooper pairs in the stable, inhomogeneous SC state are not fixed. Note that
there is a discontinuity for $|\mathbf{q|}$ at $V=V^{\ast }$, which is again
the signature of the first-order phase transition. 
\begin{figure}[tb]
\vspace{-0.85cm}
\begin{center}
\leavevmode\includegraphics[width=0.90\linewidth,clip]{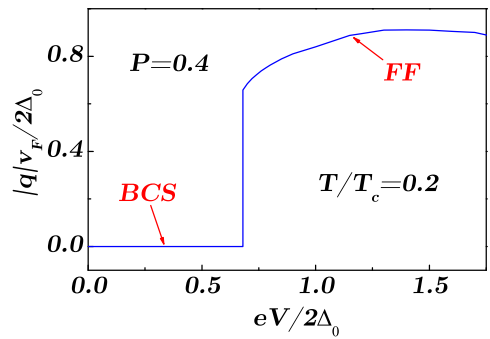}
\end{center}
\vspace{-1.0cm}
\caption{(Color online) The bias dependence of the momentum of a Cooper pair
in the antiparallel alignment of magnetizations at $P=0.4$, $T/T_{c}=0.2$,
where a discontinuity is observed.}
\label{fig2}
\end{figure}

\textit{Phase Diagram.} ---In the antiparallel configuration, a schematic
phase diagram for the central superconductor in the $T-V$ plane could be
obtained, as shown in Fig. 3 for $P=0.4$. There exist three phases: the
homogeneous BCS phase with $\Delta $ constant ($\mathbf{q}=0$) at low bias;
the inhomogeneous FF phase with $\mathbf{q}\neq 0$ for $V^{\ast }\leqslant
V<V_{c}$ at low temperature; the normal phase for $V\geqslant V_{c}$. Along
the phase boundary from B to C, the homogeneous BCS phase coexists with the
spatially modulated FF phase; along the $V_{c}$ boundary where the SC gap
closes, there are two transitions: one is from the homogeneous BCS state to
the normal state, and the other is from the spatially modulated FF state to
the normal state, which are of second-order. At point C, three phases meet,
implying it can be viewed as a \textit{Lifshitz point}\cite{lifshitz};  at
point QCP where $V_{QCP}$ satisfies $PeV_{QCP}/2\Delta _{0}=0.754$ $[=\delta
\mu (V_{QCP})]$, a quantum phase transition (QPT) from the ordered FF state
to the disordered normal state occurs at $T=0$.  
\begin{figure}[tb]
\vspace{-0.85cm}
\begin{center}
\leavevmode\includegraphics[width=0.90\linewidth,clip]{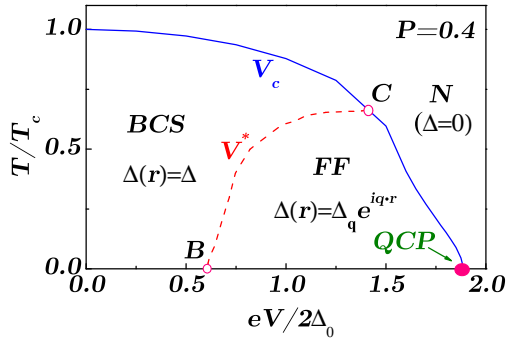}
\end{center}
\vspace{-1.0cm}
\caption{(Color online) A schematic phase diagram of the central
superconductor in the T-V plane for the antiparallel alignment of
magnetizations at $P=0.4$. Three phases are observed: the homogeneous BCS
state is separated by the $V^{\ast }$ boundary line via a first-order phase
transition from the inhomogeneous FF state, while the SC states are
separated by the $V_{c}$ boundary line via a second-order phase transition
from the normal state. A quantum phase transition is sepcified at $V_{QCP}$
(see context).}
\label{fig3}
\end{figure}

\textit{TMR.} ---The total currents are given by $I^{F}=I_{0}N^{F}$, and $%
I^{A}=I_{0}(1-P^{2})N^{A}$, where $I_{0}=2\pi e\left\vert \widetilde{T}%
\right\vert ^{2}(D_{j\uparrow }+D_{j\downarrow })$, and $N^{F(A)}$ is given
by Eq. (\ref{num-1}). The differential conductance $G^{F(A)}$ and the
tunneling magneoresistance $TMR$\ are obtained by $G^{F(A)}=\frac{dI^{F(A)}}{%
dV}$ and $TMR=\frac{G^{F}}{G^{A}}-1$.\ The bias dependence of the tunnel
conductance and the $TMR$ is shown in Fig. 4. The behavior of $G^{F}$ is
consistent with that of Ref.\cite{takahashi}, but $G^{A}$ and $TMR$ differ
from those in Ref.\cite{takahashi} owing to the intervention of the FF
state, where the oscillating behaviors for $G^{A}$ and $TMR$ are seen. These
oscillation characteristics could be served as good tests for experimentally
observing the inhomogeneous FF state, and are qualitatively consistent with
the recent observation\cite{johansson}. 
\begin{figure}[tb]
\vspace{-0.85cm}
\begin{center}
\leavevmode\includegraphics[width=0.90\linewidth,clip]{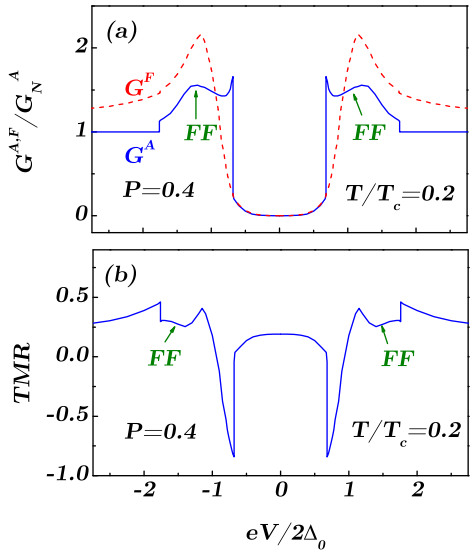}
\end{center}
\vspace{-1.0cm}
\caption{(Color online) The bias dependence of the tunnel condanctance $%
G^{F,A}$ (a) and the magnetoresistance (b) at $P=0.4$, $T/T_{c}=0.2$, where $%
G_{N}^{A}$ is the conductance in the normal state for the antiparallel
alignment. The oscillation characteristics are observed.}
\label{fig4}
\end{figure}

\textit{Quantum Criticality.} ---Since a phase transition from the
inhomogeneous FF state to the normal state appears at $V_{QCP}$, which just
happens at $T=0$, it is nothing but a manifestation of the QPT\cite{sachdev}%
. At $V\lesssim V_{QCP}$, the system is in an ordered ground state; while $%
V>V_{QCP}$, it is in a disordered ground state, suggesting $V_{QCP}$ is the
quantum critical point, where the SC gap vanishes. As $V$ approaches $%
V_{QCP} $, the amplitude of SC order parameter, the chemical potential shift
induced by the spin accumulation, as well as the difference of free energies
between the SC and normal states vanish as $\Delta _{\mathbf{q}}/\Delta _{0}$%
, $[\delta \mu (V_{QCP})-\delta \mu (V)]$, $F_{S}^{A}-F_{N}^{A}\sim
|V-V_{QCP}|^{z\nu }$ with the critical exponents $z\nu =1/2$, $1$ and $2$.
Obviously, such a QPT is of second-order. The present system offers a nice
example of the QPT.

In summary, we have revisited the spin-dependent transport in F/S/F double
tunnel junctions by taking the proximity effect into account. It is found
that in the antiparallel configuration, the first- and second-order phase
transitions, the inhomogeneous FF state, and quantum criticality can be
revealed simultaneously in the central superconductor. The present study
rectifies the previous result\cite{takahashi} where the FF state was simply
ignored.

This work is supported in part by the NSFC (Grant Nos. 90103023, 10104015,
10247002).


\begin{references}
\bibitem[\mbox{}]{Permanent address.}$^{\ast}$Corresponding author.
E-mail: gsu@gscas.ac.cn.

\bibitem{meser}  R. Meservey and P.M. Tedrow, Phys. Rep. {\bf %
238}, 173 (1994).

\bibitem{gijs}  M.A.M. Gijs and G.E.W. Bauer, Adv. Phys. {\bf 46}, 285
(1997).

\bibitem{prinz}  G. Prinz, Science {\bf 282}, 1660 (1998).

\bibitem{mood}  J.S. Moodera {\it et al.}, Annu. Rev. Mater.
Sci. {\bf 29}, 381 (1999).

\bibitem{wolf}  S.A. Wolf {\it et al.}, Science {\bf 294},
1488 (2001).

\bibitem{sarma} S. Das Sarma, American Scientist \textbf{89}, 516
(2001).

\bibitem{book} S. Maekawa and T. Shinjo (eds.), \textit{Spin-Dependent
Transport in Magnetic Nanostructures} (Taylor \& Francis, London and New
York, 2002).

\bibitem{dejong}  M.J.M. de Jong {\it et al.}, Phys. Rev. Lett. {\bf %
74}, 1657 (1995).


\bibitem{takahashi}  S. Takahashi {\it et al.}, Phys. Rev.
Lett. {\bf 82}, 3911 (1999).

\bibitem{jedema} F. J. Jedema {\it et al.},
Phys. Rev. B {\bf 60}, 16549 (1999).

\bibitem{zheng}  Z. Zheng {\it et al.}, Phys. Rev. B {\bf 62}%
, 14326 (2000).

\bibitem{tserko}  Y. Tserkovnyak {\it et al.}, Phye. Rev. B {\bf 65},
094517 (2002).

\bibitem{yoshida}  N. Yoshida {\it et al.}, Phys.
Rev. B {\bf 63}, 024502 (2000); N. Yoshida {\it et al.},
Physica C {\bf 367}, 185 (2002).

\bibitem{jin}  B. Jin, G. Su, Q. R. Zheng, M. Suzuki, Phys.
Rev. B {\bf 68}, 144504 (2003).

\bibitem{vasko}  V.A. Vasko {\it et al.}, Phys. Rev. Lett. 78, 1134 (1997); Z.W.
Dong {\it et al.}, Appl. Phys. Lett. {\bf 71}, 1718 (1997); S.K. Upadhyay {\it et al.},
Phys. Rev. Lett. {\bf 81}, 3247 (1998); N.C. Yeh {\it et al.}, Phys. Rev. B {\bf 60}%
, 10522 (1999); A. Sawa {\it et al.}, Physica C {\bf 339}, 287 (2000).

\bibitem{soulen}  R.J. Soulen {\it et al.}, Science {\bf 283}, 85
(1998). 

\bibitem{kontos} T. Kontos {\it et al.}, Phys. Rev. Lett. {\bf 86}, 
304 (2001).

\bibitem{chen}  C.D. Chen {\it et al.}, Phys.
Rev. Lett. {\bf 88}, 047004 (2002); J. Magn. Magn. Mat. {\bf 239}, 141
(2002).

\bibitem{gu} J. Y. Gu {\it et al.}, Phys. Rev. B {\bf 66}, 140507(R) (2002).

\bibitem{johansson}J. Johansson {\it et al.}, J. Appl. Phys. 93, 8650 (2003).

\bibitem{johnson}  M. Johnson {\it et al.}, Phys. Rev. Lett. {\bf 55},
1790 (1985); M. Johnson, {\it ibid}. {\bf 70}, 2142 (1993); Science {\bf 260}%
, 320 (1993); Appl. Phys. Lett. {\bf 65}, 1460 (1994).

\bibitem{FF} P. Fulde and A. Ferrel, Phys. Rev. {\bf 135}, A550 (1964).

\bibitem{LO}A. Larkin and Y. Ovchinnikov, Sov. Phys. JETP {\bf 20}, 762 (1965).

\bibitem{casal}R. Casalbuoni {\it et al.}, Rev. Mod. Phys. {\bf 76},
263 (2004) and references therein.

\bibitem{abrikosov}A.A. Abrikosov, {\it Fundamentals of the Theory of Metals} 
(North-Holland, Amsterdam,
1988), p.518.

\bibitem{heslinga} D. R. Heslinga {\it et al.}, 
Phys. Rev. B {\bf 47}, 5157 (1993).

\bibitem{lifshitz}P. M. Chaikin, T. C. Lubensky, {\it Principles of
Condensed Matter Physics} (Cambridge University Press, Cambridge, 1995), p676.

\bibitem{sachdev}S. Sachdev, {\it Quantum Phase Transitions} 
(Cambridge University Press, Cambridge, 1999).
\end{references}
\end{document}